\documentclass[aps, prl,  twocolumn, amsmath, amssymb, superscriptaddress]{revtex4-2}

\usepackage{bbold}
\usepackage{hyperref}
\hypersetup{colorlinks=true, citecolor=blue, linkcolor=blue, urlcolor=gray}
\usepackage{graphicx}
\usepackage{bm}
\usepackage{color}
\usepackage{url}
\usepackage{tikz}
\usetikzlibrary{calc}
\usepackage{amsmath,amssymb}
\usepackage{verbatim}
\usepackage{physics}
\usepackage{xcolor}
\usepackage{mwe}
\usepackage{tikz}
\usetikzlibrary{calc}

\newcommand{\bea}{\begin{eqnarray}}
\newcommand{\eea}{\end{eqnarray}}

\newcommand{\be}{\begin{equation}}
\newcommand{\ee}{\end{equation}}

\newcommand{\beal}{\begin{align}}
\newcommand{\eeal}{\end{align}}

\newcommand{\ch}{\mathcal{H}}

\newcommand{\x}{100}
\newcommand{\y}{120}
\newcommand{\z}{001}

\begin{document}


\title{Projected altermagnetism by symmetry reduction at surfaces and in thin films}

\author{Sopheak Sorn}
\affiliation{Institute for Quantum Materials and Technologies, Karlsruhe Institute of Technology, 76131 Karlsruhe, Germany}
\affiliation{Institute of Theoretical Solid State Physics, Karlsruhe Institute of Technology, 76131 Karlsruhe, Germany}
\thanks{Corresponding author: sopheak.sorn@kit.edu}

\author{Charanpreet Singh}
\affiliation{Physikalsisches Institut, Karlsruhe Institute of Technology, Karlsruhe, Germany}

\author{Lukasz Plucinski}
\affiliation{Peter Gr\"unberg Institut (PGI-6), Forschungszentrum J\"ulich GmbH, J\"ulich, Germany}

\author{Gustav Bihlmayer}
\affiliation{Peter Gr\"unberg Institut (PGI-1), Forschungszentrum J\"ulich GmbH, J\"ulich, Germany}

\author{Yuriy Mokrousov}
\affiliation{Peter Gr\"unberg Institut (PGI-1), Forschungszentrum J\"ulich GmbH, J\"ulich, Germany}
\affiliation{Institute of Physics, Johannes Gutenberg-University Mainz, Mainz, Germany}

\author{Wulf Wulfhekel}
\affiliation{Institute for Quantum Materials and Technologies, Karlsruhe Institute of Technology, 76131 Karlsruhe, Germany}
\affiliation{Physikalsisches Institut, Karlsruhe Institute of Technology, Karlsruhe, Germany}
\begin{abstract}
{
Altermagnets are a newly identified class of magnetic materials that combine vanishing net magnetization within the unit cell with spin-split electronic states. Their theoretical description relies on symmetry properties of the bulk band structure. Surfaces and thin films, however, inherently break these symmetries. Here, we investigate the consequences of such symmetry reduction for the electronic structure of bulk altermagnets near the surface and of thin films. When the surface coincides with a symmetry plane of the bulk altermagnetic order, the resulting two-dimensional Brillouin zone exhibits spin-degenerate bands, corresponding to conventional antiferromagnetic behavior. In all other cases, the symmetry of the altermagnetic order is reduced, leading to modified spin splitting. Remarkably, we discover a thin-film geometry of a $g$-wave altermagnet with a particular surface orientation that enables a $d$-wave spin splitting, which is commonly accompanied by the spin-splitter effect, suggesting the functionalization of non-$d$-wave altermagnets by surfaces.
Our findings demonstrate that symmetry breaking at surfaces and in thin films fundamentally reshapes altermagnetic spin textures, providing a tunable platform for controlling spin-dependent electronic phenomena.
}
\end{abstract}
\date{\today}

\maketitle

{Altermagnets (AMs) have recently emerged as a novel class of magnetic materials characterized by vanishing net magnetization alongside momentum-dependent spin-split electronic states dictated by crystal symmetry~\cite{LiborPRX,LiborPRX_pers}. While their response properties such as exotic transport phenomena~\cite{Takahashi_elesto-hall,Libor_Halleffect_Ruo2,Halleffect_Attias,Hall_thermal_Hoyer,Hall_Mn5Si3,wWanxiang,Fang2024,Antonenko2025MirrorChern,PRB180401} are governed by symmetry arguments applied to the three-dimensional band structure, realistic systems such as surfaces and thin films inevitably break these symmetries. This raises an important question: how robust are hallmark manifestations of AM order and sheer observability of altermagnetism in realistic samples? In particular, it is clear that surface-induced symmetry reduction can fundamentally modify the spin structure of the electronic states, implying that surface-sensitive probes may fail to detect the characteristic flavor of spin splitting expected in the bulk band structure. }

{At the same time, one can argue that the symmetry lowering upon dimensionality reduction in bulk AMs is not necessarily destructive but can be harnessed as a tool. For example, one may ask whether by engineering thin-film geometries it is possible to realize alternative spin-splitting symmetries, including the highly sought-after $d$-wave spin splitting that is accompanied by the charge-to-spin conversion functionalities \cite{Naka2019, Gonzalez2021, Shao2021, LiborPRX_pers}, even in altermagnetic materials where it is absent in the bulk. If so, this will highlight the crucial role of reduced dimensionality and symmetry breaking in shaping the electronic structure of AMs and motivate a systematic investigation of their behavior in slab geometries and at surfaces. This is further motivated by the impact that the surface-bulk symmetry mismatch may have on the claims of observability of a specific flavor of altermagnetism with surface sensitive probes.
}

\begin{figure}
    \centering
    \includegraphics[width=0.9\linewidth]{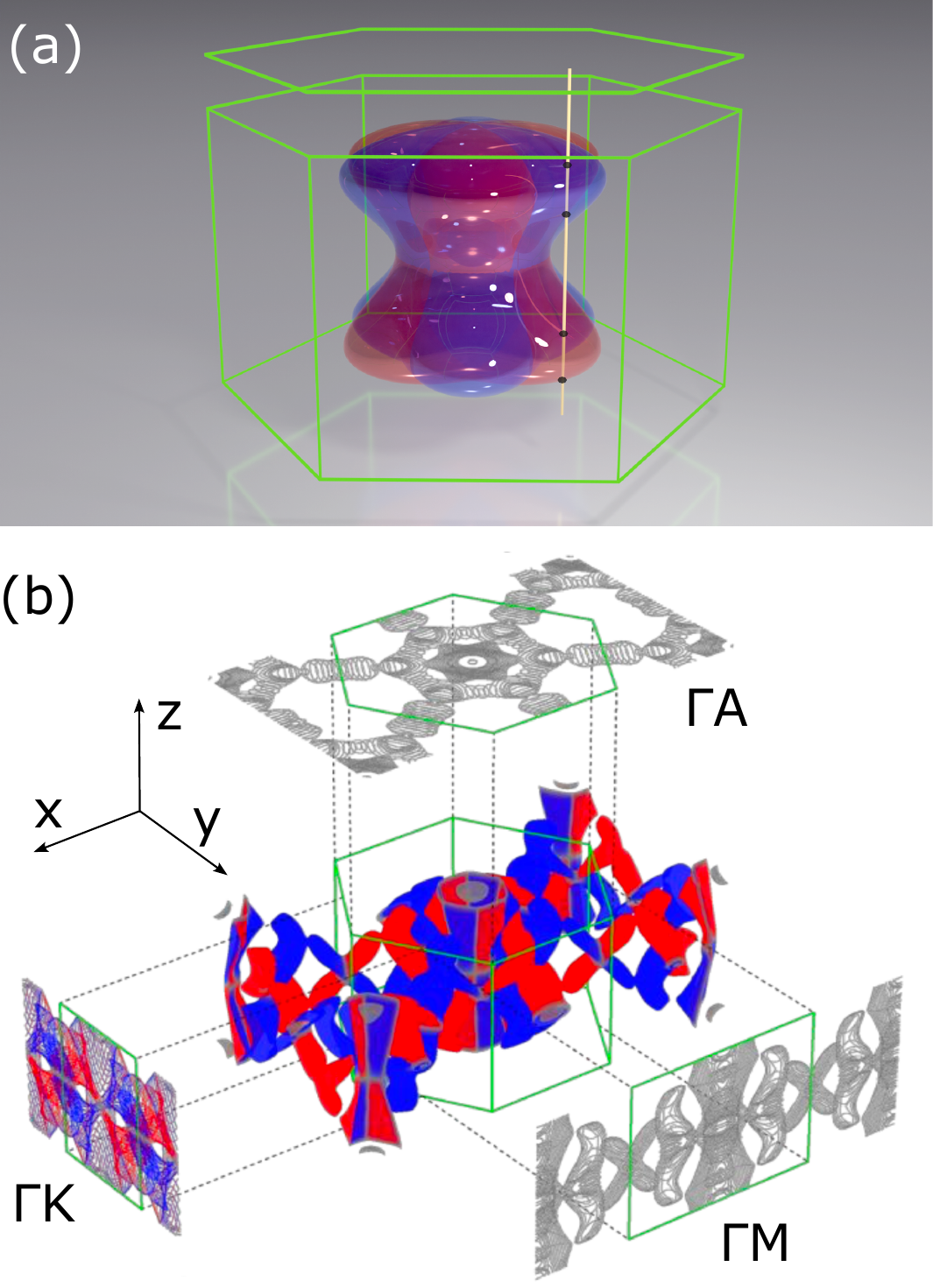}
    \caption{Illustration of bulk states projecting onto surface BZ in $g$-wave AM CrSb. (a) Schematic representation of the bulk states at the Fermi level and their $S_z$ spin polarization (blue/red) projecting onto the (001) surface BZ (topmost green hexagon) along the direction of the orange line. Blue and red bulk states project onto the same points, leading to spin-degenerate states throughout the surface BZ. (b) Projections of bulk states as computed from DFT onto three surfaces indicate a rich variety of outcomes. $\Gamma A$  and $\Gamma M$  projections yield spin-degenerate outcomes (colorless), while $\Gamma K$  projection strikingly leads to a $d$-wave spin splitting.  
    }
    \label{fig:projected}
\end{figure}

In this work, we examine the consequences of symmetry breaking at the surfaces and in thin films of AMs, using the $g$-wave AM CrSb as a representative example \cite{Reimers2024,Ding2024,Zeng_advs.202406529,Li2025,Yang2025}.
Figure \ref{fig:projected}(a) schematically shows the symmetry of the Fermi surface (or any other iso energy surface), highlighting the spin-splitting in reciprocal space. Notably, a rotation by 60$^\circ$ around the [\z ] axis maps the bands of opposite spin polarization into each other. The spin splitting vanishes on the horizontal nodal plane ($k_z = 0$) and on three vertical nodal planes. 
Moreover, the splitting has to vanish at the boundaries of the first Brillouin zone (BZ). This characterizes AMs of $g$-type.
Through a combination of group-theoretical analysis and band-structure calculations, we uncover three notable properties in the limit of vanishing spin-orbit coupling (SOC). First, for surfaces parallel to the nodal planes, we find that spin splitting in the electronic states near the surface can disappear completely. Secondly, we discover a remarkable feature for the (2$\overline{1}$0) surface: the spin splitting exhibits the highly desirable $d$-wave symmetry. Finally, for any other surface orientations, the spins on the surface plane are not fully compensated, thereby inducing a predominantly ferromagnetic-like spin-splitting feature. We elucidate how these surface phenomena are inherited from the interplay between the bulk altermagnetic order parameter (OP) and the crystalline symmetry of the surface.
Our work elucidates rich and versatile surface phenomena in AMs, calling for future investigations to examine functionalization by surface engineering.

\noindent\textit{Qualitative consideration --} 
Before scrutinizing the surface effects, we illustrate them in a semi-infinite altermagnetic system with a (\z ) surface through an over-simplified projection of bulk states onto the surface Brillouin zone;
see \footnote{Note that we follow the standard definition of the hexagonal unit cell with the basis vectors in the basal plane having an angle of 120$^\circ$. This implies that the [100] and the [110] direction form an angle of 60$^\circ$ and are equivalent. We place the cartesian $x$-axis along [\x ], the $y$-axis along [\y ] and the $z$-axis along [\z ] and use Miller indexes of crystallographic planes to indicate surface orientations.} for the surface-labeling convention. At the surface, the electrons are essentially reflected by the work function barrier leading to scattering. For a bulk Bloch state carrying the crystal momentum $\mathbf{k}=k_{\perp}\mathbf{e}_{\perp} + \mathbf{k}_{\parallel}$,
with the normal vector $\mathbf{e}_{\perp}$, its surface-parallel component $\mathbf{k}_{\parallel}$ remains a good quantum number, while the $k_{\perp}$ component is no longer conserved due to the breaking of translation symmetry at the surface.
This mechanism is illustrated in Fig.\ref{fig:projected}(a): the the reciprocal surface rods (orange line) at fixed $\mathbf{k}_{\parallel}$ project the bulk states on the surface BZ 
corresponding to the topmost hexagon of Fig.\ref{fig:projected}(a).
The scattering states consist of superpositions of incoming and outgoing bulk waves, according to their group velocity.
In the limit of vanishing SOC, spin-$S_z$ eigenvalue is also a good quantum number, so states with the different spins (red and blue) do not mix and cannot form superpositions.
As a result, the fact that (\z ) plane is a nodal plane in the bulk has a remarkable consequence. Since the plane mirrors spin-polarized bands of opposite spin character, the scattering states come in pairs with the same $k_\parallel$ momenta but opposite spin, and the scattering states of the half infinite system become spin degenerate in the whole surface BZ. 
This situation is reminiscent of the Kramers degeneracy in a conventional conllinear antiferromagnet featuring PT symmetry, yet in our case, such a symmetry is not presence, and the degeneracy arises from projecting the bulk $g$-wave bands.

The above consideration suggests a spin-polarization pattern induced by surfaces, which differs qualitatively from that of the bulk states. In addition, as it will turn out, the outcome depends strongly on the orientation of the surface, which we consider next.
The outcome of bulk projection for the density functional theory (DFT) based bulk band structure is illustrated in Fig.\ref{fig:projected}(b), where absence of spin polarization in the projected bulk states along the $\Gamma A$, i.e. $z$-direction is depicted in the colorless degenerate lines. Similarly, the projection along $\Gamma M$, i.e. $y$-direction and its symmetry equivalence are also onto nodal planes, so the same outcome is obtained.  For projections onto non-nodal planes, the spin degeneracy is generally lifted. For example, projection along $\Gamma K$, i.e. $x$-direction in Fig.\ref{fig:projected}(b) exhibits pronounced momentum-dependent spin polarization. Notably, the resulting spin texture displays a $d$-wave symmetry.
 
\begin{figure*}[bht]
    \centering
    \includegraphics[width=0.65\linewidth]
    {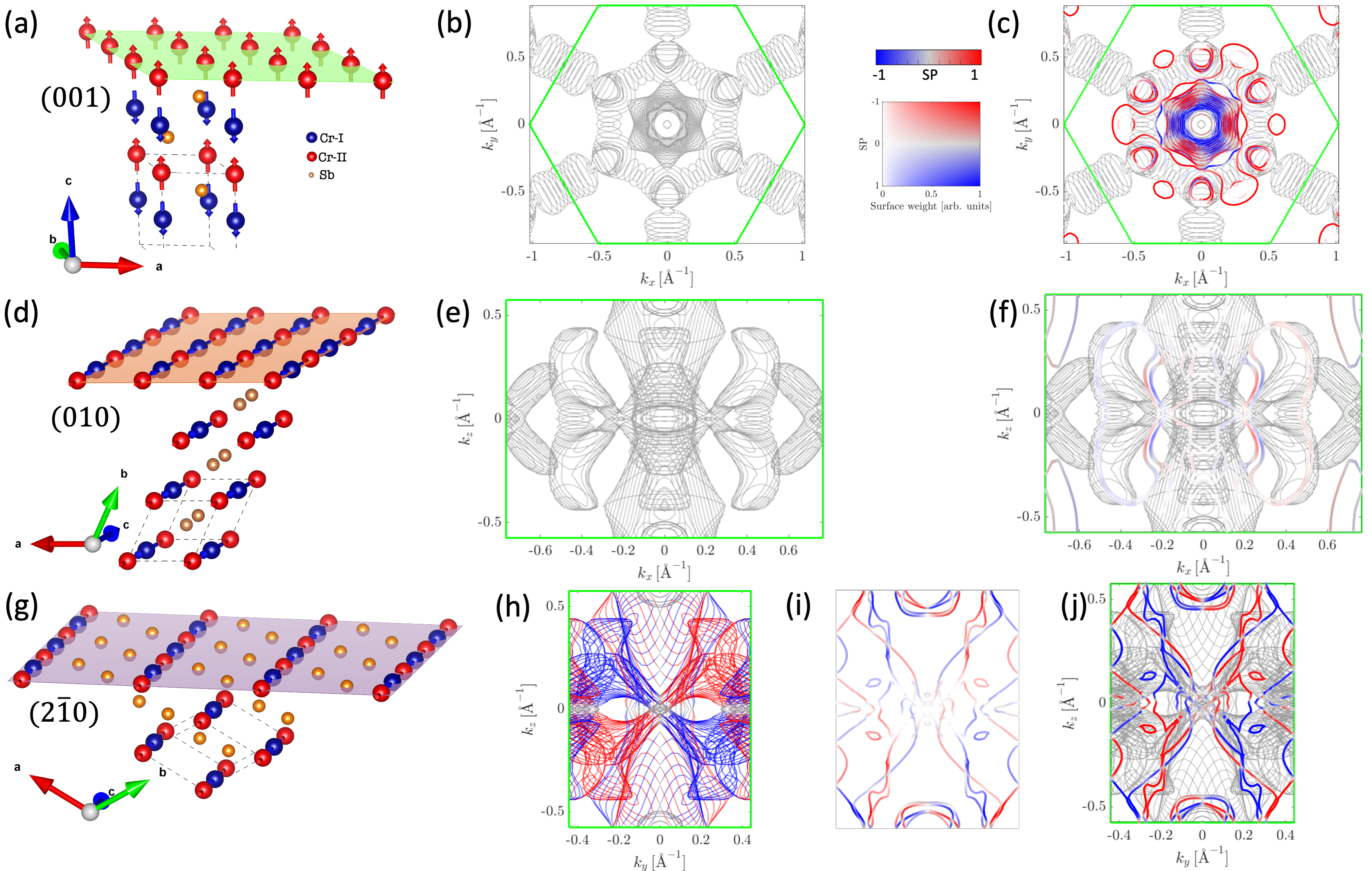}
    \caption{ Crystal structure (a,d,g) of CrSb thin slabs of (001), (010), and (2$\overline{1}$0) orientation, respectively. (b, e, h) display the color-coded $z$-spin polarization of the projected bulk states at the Fermi level onto the BZ of the slabs highlighting its absence for the (001) and (010) orientation, and its $d$-wave symmetry for the (2$\overline{1}$0) orientation. To single out surface states, we contrast them with results from the slab calculations shown in (c, f, j), through which the surface states are clearly seen as the additional lines on top of the bulk states. They exhibit pronounced $z$-spin polarization as reflected in their strong color intensity that overshadows the gray bulk states. Particularly for the (2$\overline{1}$0) plane, we remove the bulk states from (j), arriving at the surface states in (i).  
    All first principle calculations were performed with SOC included.   
    }
    \label{fig:SlabZ}
\end{figure*}

Our consideration points to a rich variety of surface phenomena in AMs, where the momentum dependence of the spin splitting at the surface can differ qualitatively from that in the bulk.
However, our current discussion is based on a simplified bulk-projection picture and does not yet incorporate microscopic surface details, which may qualitatively modify these conclusions.
Indeed, the dipole moments at the (001) plane, e.g. the outermost layer of spins in Fig.\ref{fig:SlabZ}(a), are uncompensated. As shown later, this reinstates a spin splitting resembling that in a ferromagnet. On the contrary, the dipole moments on the (010) and (2$\overline{1}$0) surfaces are compensated (neglecting SOC corrections), so the qualitative conclusions of the bulk-projection picture continue to hold. 

\noindent \textit{Group-theoretical analysis}---
To substantiate the preceding simple picture, we provide evidence from a group-theory analysis \cite{gtpack1, gtpack2}, through which we establish connection between the bulk altermagnetic OP with the spin-splitting properties near surfaces. We will examine symmetry-allowed terms in the Bloch Hamiltonian induced by the OP in the presence of a surface, whose form resembles a Zeeman term with a coefficient that depends on $\mathbf{k}_{\parallel}$. It is this coefficient that encodes the information about the momentum-dependent spin splitting of the electronic states.

Generally, an AM supports a ferroic ordering of magnetic multipole moments (MMM) in its bulk, which can be treated as its OP \cite{Steward2023, Spaldin2024, McClarty2024, Schiff2024}
\bea
    \mathcal{O}_{i_1 i_2 \cdots i_{m-1}, i_m} &=& \int d^dr\ r_{i_1}\cdots r_{i_{m-1}} \mu_{i_m}(\boldsymbol{r}),
    \label{eq:mmm}
\eea
where $\boldsymbol{\mu}$ is the local magnetization density, $i_1, \cdots, i_{m-1}$ are spatial indices, while $d$ is the spatial dimension. For simplicity, we will focus on $d=3$, and the generalization to $d=2$ is straightforward. 
Landau theory of AMs has uncovered two important facts in the absence of SOC \cite{McClarty2024, Schiff2024}: (1) $m=3$ is the lowest possible rank in AMs, corresponding to magnetic octupole moments, which are stabilized by non-relativistic exchange interactions and the lattice effects, and (2) for fixed $i_1, \dots, i_{m-1}$, the vector $\boldsymbol{\mathcal{O}}_{i_1\cdots,i_{m-1}}$ has its orientation determined by the N\'eel vector $\boldsymbol{N}$, i.e., they are collinear. In the crystalline environment, symmetry allows only certain linear combinations of $\mathcal{O}_{i_1\cdots i_m}$ to be nonzero. SOC generally produces easy axes that prefer $\boldsymbol{N}$ to point along certain high-symmetry directions in the crystal. In addition, depending on the direction of $\boldsymbol{N} $, a magnetic dipole moment ($m = 1$) can be generated by SOC, i.e. weak ferromagnetism \cite{PhysRevLett.134.196703,Autieri2025}. 
In the rest of the manuscript, we assume that such weak ferromagnetism, if present, is small so that the spin splitting (SS) in the band structure is predominantly governed by the MMMs associated with the non-relativistic exchange effects.

\begin{table}[t]
    \centering
    \begin{tabular}{l l l l}
    \hline
    Surface & Nonrelativistic & SOC-enabled\\
            & spin splitting  & spin splitting\\
    \hline
    \hline
    (001)   & $\varphi \sigma_z$ & $\lambda_R(k_x\sigma_y - k_y \sigma_x) + $\\
    \vspace{5 pt}
            &                    & $\varphi \left[2k_xk_y \sigma_x + (k_x^2 - k_y^2) \sigma_y \right]$ \\
    \hline
    \vspace{5 pt}
    (010)   & 0                  & $\left(\lambda_{R,x} k_z\sigma_x - \lambda_{R,z} k_x \sigma_z \right) + \varphi \sigma_y$\\
    \hline
    \vspace{5 pt}
    (2$\overline{1}$0)   & $\varphi k_yk_z \sigma_z$ & $\left(\lambda_{R,z} k_y\sigma_z - \lambda_{R,y} k_z \sigma_y \right) + \varphi \sigma_y$\\
    \hline
    \end{tabular}
    \caption{Group-theoretical analysis of spin splitting in CrSb due to surfaces. Symmetry-allowed terms in the Bloch Hamiltonian involve Pauli matrices corresponding to electronic spin operators. The terms in the second column dominate in the small-SOC limit and can be regarded as Zeeman terms with momentum-dependent coefficients which govern the spin splitting behaviors near the surfaces.}
    \label{tab:symmetry1}
\end{table}

Bulk AM CrSb with the N\'eel vector along [001] axis supports a rank-5 MMM as its OP $\varphi$
\bea
\varphi \equiv \mathcal{O}_{(3x^2-y^2)yz,z} = 3\mathcal{O}_{xxyz,z} - \mathcal{O}_{yyyz,z}.
\eea
It is responsible for the $g$-wave SS in the bulk bands with the $z$-spin polarization and the form factor $F(\boldsymbol{k}) = (3k_x^2 - k_y^2)k_yk_z$ at small $|\boldsymbol{k}|$. This can be seen through the low-energy analysis where the lattice point group $D_{6h}$ allows for the SS term in the electron Bloch Hamiltonian which assumes the following form near the $\Gamma$ point in the BZ: $\ch_{\rm SS} \propto \varphi F(\boldsymbol{k})\sigma_z$; see Ref.\cite{Fernandes2024} for detailed procedures. As mentioned before, such a term acts like a Zeeman field that has a momentum dependence as given by the form factor $F(\boldsymbol{k})$. Below, we follow the same procedure to obtain the SS structure at surfaces.

For the (001) surface, the lattice point group is reduced to $C_{3v}$, and the surface crystal momentum in this case is denoted by $\mathbf{k}_{\parallel}=(k_x, k_y)^T$. The bulk OP $\varphi$ transforms as a time-reversal-odd $A_2^-$ irreducible representation of $C_{3v}$.  Our analysis shows that the leading SS term in the Bloch Hamiltonian induced by $\varphi$ is given by $\mathcal{H}_{SS}^{(001)} \propto \varphi \sigma_z$ near the $\Gamma$ point. 
As it turns out, the form factor in the leading term is independent of $\mathbf{k}_{\parallel}$, meaning that the electronic states near the surface exhibit a spin-split structure that is predominantly $\mathbf{k}_{\parallel}$-independent, just like a ferromagnet. This is consistent with our qualitative consideration at the beginning, where this outcome stems from the uncompensated spin moments on the (001) surface. We remark that we have purposely kept only the symmetry-allowed terms that involve $\sigma_z$ since they arise microscopically from the nonrelativistic exchange field due to the N\'eel vector along the z-axis. 
Terms involving $\sigma_x$ and $\sigma_y$ are only possible in the presence of SOC, which we consider next. The leading terms in the Taylor expansion in $\mathbf{k}_{\parallel}$ is $\mathcal{H}_{SS, SOC}^{(001)} \propto \lambda_R \left(k_x \sigma_y - k_y \sigma_x\right) + \varphi\left[2k_xk_y \sigma_x + (k_x^2 - k_y^2)\sigma_y\right]$. The first contribution is the Rasha term, which exists even without the altermagnetic order thank to the broken inversion symmetry at the surface, while the second term is due to the altermagnetic order in the bulk. In the small-SOC limit, the spin splitting is still dominated by the nonrelativistic term $\varphi\sigma_z$, whereas the relativistic terms stand as small corrections.

Following the same procedure, we obtain results for (010) and (2$\overline{1}$0) surface, as summarized in Table \ref{tab:symmetry1}. We omit the computation steps and go straight to the discussion of the results. For the (010) surface, the reduced point group $C_{2v}$ forbids any coupling between $\varphi$ and $\sigma_z$, resulting in spin degenerate bands, resembling the Kramers degeneracy in a $PT$-symmetric antiferromagnet; see Supplementary Materials (SM) for a detailed explanation. Interestingly, this agrees with outcomes from the bulk-band projection along $\Gamma M$ in Fig.\ref{fig:SlabZ}(e). SOC leads to small spin splitting whose spin polarization and momentum dependence are given by $\mathcal{H}_{SS, SOC}^{(010)}\propto (\lambda_{R,x}k_z \sigma_x - \lambda_{R, z}k_x \sigma_z)+\varphi \sigma_y$. The first contribution in the parentheses is a modified Rashba term, while the second contribution $\varphi\sigma_y$ is equivalent to a ferromagnetic splitting. The latter indicates the rise of a surface weak ferromagnetism.

Finally, for the (2$\overline{1}$0) surface, our analysis shows a nonrelativistic spin splitting with a $d$-wave symmetry, characterized by $\mathcal{H}_{SS}^{(2\overline{1}0)} \propto \varphi k_yk_z\sigma_z.$ In the presence of SOC, correction terms similar to those for the (010) surface are obtained, including Rashba-like terms and weak ferromagnetism. In the Supplementary Material, we apply the same procedure for slab geometry in the orientations as these surfaces, highlighting the comparison between the two opposite surfaces of each slab geometry. We also analyze surface spin splitting for MnTe \cite{Lee2024,Krempasky2024,PhysRevB.109.115102} whose essential distinction from CrSb is the direction of the N\'eel vector.

\noindent \textit{Electronic structure from ab initio}---We now support our group-theory results by performing explicit first principles calculations of thin films of CrSb including the SOC effect and keeping the direction of the N\'eel vector along the bulk [001]-axis ($z$-axis) in all cases, with computational details and complementary tight binding calculations \cite{PhysRevB.110.144412} presented in the Supplementary Material.
To extract the surface effects, we present results comparing side-by-side between (a) bulk band structure projected on surface BZ and (b) band structures obtained from finite-slab computations.

Figure \ref{fig:SlabZ}(b, e, h) shows the projection of bulk states at the Fermi level onto the surface BZ for the (001), (010), and (2$\overline{1}$0) plane, respectively. We observe that only the projection on the (2$\overline{1}$0) plane exhibits a noticeable spin splitting carrying the $d$-wave symmetry in the surface BZ, see Fig.\ref{fig:SlabZ}(h), while the other two orientations feature a spin-degenerate projection, Fig.\ref{fig:SlabZ}(b,e), even with SOC. As mentioned before, the detailed surface properties will modify these outcomes. 

To see the surface effects in each orientation, we compare its bulk projection with its electronic structure in the slab geometry. The latter is shown in Fig.\ref{fig:SlabZ}(c,f,j) for the (001), (010), and (2$\overline{1}$0) surface, respectively. The thickness and the color of the lines encode the information about the weight and the $S_z$ spin polarization, respectively, of the eigenstates on the outermost chromium sites at the surface. For the (001) film, we observe clear signatures of spin polarization. By identifying states with large weight and by comparing with the bulk projection in panel (b), we can observe extra surface states that form loop-like structures midway along the $\Gamma K$ paths. Near the BZ center, the spin polarization features an $s$-wave symmetry, consistent with the group-theory results, which we attribute to the uncompensated dipole moments on the outermost plane of Cr sites. We have checked that such a spin polarization occurs even without SOC.
For larger $|\mathbf{k}_{\parallel}|$ away from the BZ center, the spin polarization acquires a momentum dependence that reflects a three-fold rotation symmetry about the [001] axis. In principle, this can be understood by keeping higher-order terms in $|\mathbf{k}_{\parallel}|$ in the group-theory analysis.

While the (001) surface supports $z$-spin polarization already without SOC, this is not the case for the (010) surface in Fig.\ref{fig:SlabZ}(d). 
Like before, the surface states can be distinguished from the bulk states by contrasting panel (e) and (f). 
As it turns out, the $z$-spin polarization of the surface states emerges only when SOC is included into the calculation. Its momentum dependency appearing to have a $p$-wave symmetry can be explained by the Rashba-like term of the form $k_x\sigma_z$; see Table \ref{tab:symmetry1}.

Figure \ref{fig:SlabZ}(h) shows the electronic states at the Fermi level in a (2$\overline{1}$0) slab geometry. To facilitate our discussion, we isolate the surface states and show them in panel (i) which is obtained by artificially removing the bulk states in panel (h) from panel (j). We find that the momentum-dependent $z$-spin polarization in (i) carries predominantly a $d$-wave character, like the bulk projection in panel (h). We also identify two additional key features: (i) surface states exhibit a more pronounced spin polarization, compared with the bulk states within the slab, which is clearly seen from panel (g), where the pronounced blue/red color of the surface states eclipses that of the bulk states, and (ii) a weak asymmetry between left and right in panel (f) and (g), which can be understood from the Rasha-like SOC correction of the form $k_y \sigma_z$ reported in Table \ref{tab:symmetry1}. Remarkably, the d-wave character becomes exact when SOC is neglected. We emphasize that the key features of the slab band structure are in good overall agreement with our group-theoretical analysis.

\noindent \textit{Summary and Discussion}---
Our work demonstrates how the bulk OP of an AM governs the spin splitting of electronic states near its surface, giving rise to a variety of spin splitting whose momentum dependence differs from that of the bulk and depends sensitively on the surface orientation. In the representative example of CrSb, we find ferromagetic-like spin splitting in the electronic states near its (001) surface, while the (010) surface supports only weak spin splitting that scales with the SOC strength. Remarkably, the (2$\overline{1}$0) surface stabilizes the highly desirable $d$-wave spin splitting which is absent in the bulk.  Our findings have direct implications for spectroscopic and transport experiments on altermagnetic thin films. Surface-sensitive probes such as spin- and angle-resolved photoemission will be affected by surface states which carries a distinct spin-splitting symmetry from its bulk counterpart. Our work calls for a careful interpretation of surface-sensitive experiments.
Interestingly, projection on the surfaces that are nodal planes leads to fully degenerate bands in the surface BZ. In this situation, the total energy of the system can be lowered by electronic or structural reconstruction of the surface adding complexity to the problem~\cite{Korshunov2026CrSb}.
Meanwhile, transport experiments, which measure film-averaged properties, will receive contributions from both surfaces of a thin-film slab geometry. Our analysis in SM reveals that the $d$-wave spin splitting on both surfaces of a (2$\overline{1}$0) CrSb thin film contributes with the same sign to the charge-to-spin conversion functionality, i.e. the spin-splitter effect \cite{Naka2019, Gonzalez2021, Shao2021, LiborPRX_pers}, which is separately studied in Ref.~\cite{Sorn2026}. Therefore, our work calls for further experimental investigations into the functionalization of an AM by surface engineering.

\noindent {\it Acknowledgements}. We would like to thank Markus Garst, Iksu Jang, and Tom Saunderson for helpful discussions. 
The authors gratefully acknowledge the Deutsche Forschungsgemeinschaft,  DFG (TRR 173/3-268565370 Spin+X, project A11; TRR 288/2-422213477 Elasto-Q-Mat: project A11 and B06), the Horizon Europe Framework, grant number 101226840 (ORBIS) and grant number 101129641 (OBELIX). L.P., G.B. and Y.M. acknowledge the Jülich Supercomputing Centre for providing computational resources under projects jiff40. 

\bibliography{refs}
\end{document}



\title{Supplementary Material: Projected altermagnetism by symmetry reduction at surfaces and in thin films}

\author{Sopheak Sorn}
\affiliation{Institute for Quantum Materials and Technologies, Karlsruhe Institute of Technology, 76131 Karlsruhe, Germany}
\affiliation{Institute of Theoretical Solid State Physics, Karlsruhe Institute of Technology, 76131 Karlsruhe, Germany}

\author{Charanpreet Singh}
\affiliation{Physikalsisches Institut, Karlsruhe Institute of Technology, Karlsruhe, Germany}

\author{Lukasz Plucinski}
\affiliation{Peter Gr\"unberg Institut (PGI-6), Forschungszentrum J\"ulich GmbH, J\"ulich, Germany}

\author{Gustav Bihlmayer}
\affiliation{Peter Gr\"unberg Institut (PGI-1), Forschungszentrum J\"ulich GmbH, J\"ulich, Germany}

\author{Yuriy Mokrousov}
\affiliation{Peter Gr\"unberg Institut (PGI-1), Forschungszentrum J\"ulich GmbH, J\"ulich, Germany}
\affiliation{Institute of Physics, Johannes Gutenberg-University Mainz, Mainz, Germany}

\author{Wulf Wulfhekel}
\affiliation{Institute for Quantum Materials and Technologies, Karlsruhe Institute of Technology, 76131 Karlsruhe, Germany}
\affiliation{Physikalsisches Institut, Karlsruhe Institute of Technology, Karlsruhe, Germany}
\date{\today}

\maketitle

\beginsupplement
\section{Details of symmetry analysis for thin films of CrSb}

\subsection{(001) thin films}
The point group for the (001) CrSb films, i.e., films oriented in the xy-plane, as in Fig.2(a) is $D_{3h}$, with the surface crystal momenta $\vec k_{\parallel} \equiv \vec q = (q_x, q_y)^T$. 
The elements of the group are $\{e, C^+_{3[001]}, C^{-}_{3[001]}, \mathcal{M}_{[001]}, S^+_{6[001]}, S^{-}_{6[001]}, \mathcal{M}_{[110]},\mathcal{M}_{[100]}, \mathcal{M}_{[010]}, C_{2[1\bar{1}0]}, C_{2[120]}, C_{2[210]}\}$, where $e$ is the identity, $C^{\pm}_{n}$ is the $n-$fold rotation about a specific axis with a superscript $\pm$ denoting the counterclockwise and clockwise sense, respectively. The convention of the Miller index [uvw] for the rotation axis is used \cite{Ashcroft1976}.  
$\mathcal{M}_i$ is the mirror plane perpendicular to the $i$ axis. $S^{\pm}_n$ denotes the $n$-fold improper rotation.

Mirror symmetry $\mathcal{M}_{[001]}$, i.e., the z-mirror, prevents the magnetic multipolar order parameter $\varphi$ from coupling to the spin Pauli matrix $\sigma_z$ to form the spin splitting (SS) term in the Bloch Hamiltonian. The reason is because the bulk OP $\varphi$ is odd under the mirror, while $\sigma_z$ is even. Since none of the crystal momentum components of the surface are odd under the mirror operation, it is not possible to write down a symmetry-allowed term that involves $\varphi, \sigma_z$ and the crystal momentum. 

Equivalently, one can also see this from the fact that the SS term $\mathcal{H}^{(001)}_{\rm SS} \propto \varphi F(\vec q) \sigma_z$ does not exist for a \emph{nonzero} $F(\vec q)$. We explain this step-by-step as the following. First, $\varphi$ transforms as the time-reversal-odd one-dimensional $A_1''^-$ irreducible representation of $D_{3h}$. Second, for any scalar function $F(\vec q)$ that transforms like a one-dimensional irreducible representation for $D_{3h}$, the representation must be $A_1'^{\pm}$ or $A_{2}'^{\pm}$. Since $\sigma_z$ transforms as $A_2'^-$, then the combination $F(\vec q) \sigma_z$ transforms either like $A_1'^{\pm}$ or $A_2'^{\pm}$. These cannot combine with $A_1''^-$ (i.e. $\varphi$) to form a symmetry-allowed SS term, which must transform like $A_1'^+$. Therefore, $\mathcal{H}_{\rm SS}$ is trivial, i.e. vanishes.

In the presence of SOC, we need to analyze SS terms that involve Pauli spins that are not collinear with the N\'eel vector, i.e., $\mathcal{H}^{(001)}_{\rm SS,SOC} \propto \varphi \left[F_x(\vec q)\sigma_x + F_y(\vec q) \sigma_y\right]$. Our analysis shows that, for the lowest order of $F_{x,y}$ in $\vec q$, $\mathcal{H}_{\rm SS, SOC}^{(001)} \propto \varphi \left[2q_x q_y \sigma_x + (q_x^2 - q_y^2)\sigma_y\right]$. This term dominates the SS behavior since the non-relativistic effect is absent in the (001) films.

\begin{table*}[h]
    \centering
    \caption{Summary of symmetry analysis for thin films and surfaces of CrSb (N\'eel vector along [001]). Remarks: In the \emph{thin-film} settings considered here, the mirror symmetry in the film plane forbids the Rashba term $\mathcal{H}_{\rm SOC} \propto \hat{n}\cdot (\vec q \times \vec{\sigma})$, where $\hat n$ is the surface normal vector. In contrast, such a mirror symmetry is broken at the \emph{surfaces}, thereby allowing for the Rashba term. As a result, the SS correction term $\mathcal{H}_{\rm SS, SOC}$ becomes just a SOC contribution in addition to the Rashba term. }
    \begin{tabular}{l l l l}
        \hline
         Orientation & Point group & Nonrelativistic SS term & SOC-driven SS term\\
         & & $\mathcal{H}_{\rm SS} \propto$ &  \\ 
         \hline
         \hline
         $(001)$ thin film    & $D_{3h}$ & $0$ & $\varphi \left[2q_xq_y\sigma_x + (q_x^2-q_y^2)\sigma_y\right]$\\
         $(001)$ surface      & $C_{3v}$ & $\varphi \sigma_z$    & $\varphi \left[2q_xq_y\sigma_x + (q_x^2-q_y^2)\sigma_y\right] + \lambda_R (q_x \sigma_y - q_y \sigma_x)$\\
         \hline
         $(010)$ thin film    & $D_{2h}$ & $0$   & $\varphi \sigma_y$ \\
         $(010)$ surface      & $C_{2v}$ & $0$   & $\varphi \sigma_y + \lambda_{R,x}q_z \sigma_x - \lambda_{R,z} q_x \sigma_z$ \\
         \hline
         (2$\overline{1}$0) thin film    & $D_{2h}$ & $\varphi q_yq_z\sigma_z$  & $\varphi \sigma_y$ \\
         (2$\overline{1}$0) surface      & $C_{2v}$ & $\varphi q_yq_z\sigma_z$  & $\varphi \sigma_y + \lambda_{R,z} q_y \sigma_z - \lambda_{R,y} q_z \sigma_y$ \\
         \hline
    \end{tabular}
    \label{tab:CrSbSummary}
\end{table*}

\subsection{(010) thin films}
The point group for the (010) CrSb films, i.e., films oriented in the zx-plane, as in Fig. 2(d) is $D_{2h}$ with the surface crystal momenta $\vec k_{\parallel} \equiv \vec q = (q_x, q_z)^T$. The elements of the group are listed in the following: $\{e, \mathcal{P}, 
\mathcal{M}_{[001]}, \mathcal{M}_{[100]}, \mathcal{M}_{[120]}, C_{2[001]}, C_{2[100]}, C_{2[120]}\}$. $\mathcal{P}$ is the inversion. 
The analysis for SS is analogous to that for the case of (001) thin films. The order parameter $\varphi$ transforms as the time-reversal-odd $B_{1g}^-$ irreducible representation of $D_{2h}$.
Without SOC, the SS is absent due to mirror $\mathcal{M}_{[120]}$, analogous to the scenario in the (001) film. 
In the presence of SOC, the SS is described by the symmetry-allowed term $\mathcal{H}_{\rm SS, SOC}^{(010)} \propto \varphi \sigma_y$, where we have kept only the leading term in the expansion in $|\vec q|$. Notably, the SS generated by SOC is ferromagnetic, i.e., the weak ferromagnetism generated by SOC. Therefore, the SS in the (010) thin film is caused by the weak ferromagnetism since the nonrelativistic contribution is forbidden by symmetry. Since the weak ferromagnetic field is normal to the plane, the thin film also supports an anomalous Hall effect.

\subsection{(2$\overline{\it1}$0) thin films}
The point group for the (2$\overline{1}$0) CrSb films, i.e., films oriented in the yz-plane,  as in Fig. 2(d) is $D_{2h}$ with the surface crystal momenta $\vec k_{\parallel} \equiv \vec q = (q_y, q_z)^T$. The symmetry elements are $\{e, \mathcal{P}, \mathcal{M}_{[001]}, \mathcal{M}_{[100]}, \mathcal{M}_{[120]}, C_{2[001]}, C_{2[100]}, C_{2[120]}\}$. As described in the main text, the band structure acquires a $d$-wave SS, already in the non-relativistic limit without SOC. In the presence of SOC, the non-relativistic description acquires a correction, given by $\mathcal{H}_{\rm SS, SOC}^{(2\overline{1}0)} \propto \varphi \sigma_y$. We have kept only the leading term in the expansion in $|\vec q|$. This is analogous to the weak ferromagnetism in the (2$\overline{1}$0) case. Therefore, the $d$-wave SS in the (2$\overline{1}$0) thin film is accompanied by weak ferromagnetism.

Finally, we remark on the the charge-to-spin conversion functionality that accompanies the $d$-wave SS. This functionality is known in the literature as the spin-splitter effect\cite{LiborPRX_pers}, where a longitudinal charge current is converted into a transverse spin current, resembling the spin Hall effect, yet it is a hallmark to altermagnetism, which does not rely on the SOC strength. This effect is inherent to $d$-wave altermagnetism, which, in our case, is absent in the bulk and is enabled by the thin-film geometry. Our symmetry analysis, further substantiated by the \textit{abinitio} results, shows clear evidence that $(2\overline{1}0)$ surfaces of thin films support the spin splitter effect. In addition, for the thin-film geometry, the opposite surfaces realize the spin-splitter effect with the \emph{same} sign, leading to a constructive outcome, instead of a cancellation. Details of this effect are studied in the companion publication Ref.\cite{Sorn2026}.

\section{Details of symmetry analysis for surfaces of CrSb in semi-infinite setting}
In the semi-infinite setting, the symmetry group becomes different from the thin-film case. Nonetheless, the symmetry analysis proceeds analogously to before. 

\subsection{(001) surface}
Elements in the point group for such a surface form a subset of the corresponding thin-film point group. The half-infinite point group consists of elements of the thin-film point group that do not interchange the two surfaces of the thin film. Therefore, the elements of the (001) half-infinite point group are $\{e, C_{3[001]}^+, C_{3[001]}^-, \mathcal{M}_{[100]}, \mathcal{M}_{[010]}, \mathcal{M}_{[110]}\}$, corresponding to $C_{3v}(3m)$ point group. The order parameter $\varphi$ transforms like $A_2^-$, which leads to the symmetry-allowed $\mathcal{H}_{\rm SS}^{(001)\rm surface} \propto \varphi \sigma_z$. This term is absent in the thin-film analysis due to the fact that the analysis takes into account the combined effect of the two surfaces of the thin film. Therefore, we can conclude from both analyses that the two surfaces of the thin film realize \emph{opposite} ferromagnetic spin splitting, which combines to produce spin-degenerate bands in the thin-film setting (in the absence of SOC.) In the presence of SOC, a correction term of the form $\mathcal{H}_{\rm SS, SOC}^{(001)\rm surface} \propto \lambda_R (q_x \sigma_y - q_y \sigma_x) + \varphi \left[2q_xq_y\sigma_x + (q_x^2-q_y^2)\sigma_y\right]$ arises. As discussed in the main text, the first term is a Rashba term arising from the broken inversion at the surface, while the second term is the spin-splitting term induced by the altermagnetic order parameter $\varphi$.

Structurally, we have both a three-fold rotation axis at the (001) surface and mirror planes that reflect with respect to the $yz$-plane. In the Fermi-surface plots, projected on the surface Cr atoms, this leads to a six-fold symmetric image. In the presence of SOC, the Rashba term due to the potential gradient along the surface-normal vanishes (since the spin also points in $z$-direction), but the in-plane gradient leads to a non-vanishing term $\pm E_y \cdot (k_x \times \sigma_z)$. This adds a spin-polarization with $\sigma_R (k_x) = -\sigma_R (-k_x)$ to the spin-polarization created by the surface Cr atoms, leading to the asymmetric appearance with respect to reflection along $k_x$ seen in  Fig. 2(b). This is also compatible with the fact that the reflection at the $yz$-plane inverts the out-of-plane spin component.

\subsection{(010) surface}
By the same token as in the analysis for (001) surface, we arrive at the point group for the (010) surface $C_{2v}(mm2)$ which consists of the following elements: $\{e, \mathcal{M}_{[001]}, \mathcal{M}_{[100]}, C_{2[120]}\}$. The order parameter $\varphi$ transforms as $A_2^-$ irreducible representation of $C_{2v}$. The symmetry analysis shows that it is not possible to have a nontrivial form factor $F(q_z,q_x)$ such that $\varphi F(q_z, q_x)\sigma_z$ is a symmetry-allowed term. Therefore, there is no spin splitting due to the nonrelativistic effect. Nonetheless, in the presence of SOC, $\mathcal{H}_{\rm SS, SOC}^{(010)\rm surface} \propto \left( \lambda_{R, x} q_z \sigma_x - \lambda_{R, z} q_x \sigma_z \right) + \varphi \sigma_y$. The first term is a modified Rashba term brought about by the broken inversion symmety at the surface, whereas the second term corresponds to a weak ferromagnetism. Therefore, the spin splitting is governed by the relativistic effect where a Rashba term is superimposed with weak ferromangetism induced by $\varphi$. In comparison with the $(010)$ thin film where only the weak ferromagnetism is present, we can conclude that the two surfaces of the thin film support weak ferromagnetism with the same sign while hosting the Rashba term with the opposite signs, leading to its absence in the thin-film results. The weak ferromagnetism naturally implies the presence of anomalous Hall effect enabled by the film geometry, which is absent in the bulk.

For this surface the spin is in-plane, so the potential gradient along the surface normal ($y$) can lead to a Rashba effect in the $k_x$ direction, $\pm E_y \cdot (k_x \times \sigma_z)$. Since there is no projected spin-polarization from the bulk states and structurally there are two mirror planes that lead to vanishing in-plane potential gradients, this is the only source of spin-polarization (cf.\ Fig. 2(g)).

\subsection{(2$\overline{\it 1}$0) surface}
The point group for the (2$\overline{1}$0) surface in Fig. 2(g) is $C_{2v}$ encompassing the following elements: $\{e, \mathcal{M}_{[001]}, \mathcal{M}_{[120]}, C_{2[100]}\}$. The order parameter $\varphi$ transforms like the time-reversal-odd one-dimensional irreducible representation $B_2^-$ of $C_{2v}$, giving rise to a $d$-wave SS term $\mathcal{H}_{\rm SS}^{(2\overline{1}0) \rm surface} \propto \varphi q_yq_z\sigma_z$. SOC generates corrections of the form $\mathcal{H}_{\rm SS, SOC}^{(2\overline{1}0)\rm surface} \propto (\lambda_{R,z} q_y \sigma_z - \lambda_{R,y} q_z \sigma_y) + \varphi \sigma_y$. Therefore, similar to the thin-film setting, the $(2\overline{1}0)$ surface supports $d$-wave SS accompanied by weak ferromagnetism, while a Rashba-like term exists even before the altermagnetic order sets in.

Here, we find again a Rashba term of the form $\pm E_x \cdot (k_y \times \sigma_z)$ (note, that now $x$ is the surface normal), but it is superimposed on the projected bulk polarization. This leads to an antisymmetry in $k_y$ but an asymmetry in $k_z$ (cf.\ Fig. 2(e)).

\begin{table*}[t]
    \centering
    \caption{Summary of symmetry analysis for thin fims and surfaces of MnTe (N\'eel vector along [120]).  Remarks: In the \emph{thin-film} settings considered here, the mirror symmetry in the film plane forbids the Rashba term $\mathcal{H}_{\rm SOC} \propto \hat{n}\cdot (\vec q \times \vec{\sigma})$, where $\hat n$ is the surface normal vector. In contrast, such a mirror symmetry is broken at the \emph{surfaces}, thereby allowing for the Rashba term. As a result, the SS correction term $\mathcal{H}_{\rm SS, SOC}$ becomes just a SOC contribution in addition to the Rashba term. }
    \begin{tabular}{l l l l}
        \hline
         Orientation & Point group & Nonrelativistic SS term & SOC-driven SS correction term\\
         & & $\mathcal{H}_{\rm SS} \propto$ & $\mathcal{H}_{\rm SS, SOC} \propto$ \\ 
         \hline
         \hline
         $(001)$ thin film    & $D_{3h}$ &  0 &  $\varphi_y(3\varphi_x^2 - \varphi_y^2) \sigma_z$\\
         $(001)$ surface      & $C_{3v}$ & $\varphi_x \sigma_x +\varphi_y \sigma_y$    & $\varphi_y(3\varphi_x^2 - \varphi_y^2) \sigma_z + \lambda_R (q_x \sigma_y - q_y \sigma_x)$\\
         \hline
         $(010)$ thin film    & $D_{2h}$ &  0  &  $\varphi\sigma_z$  \\
         $(010)$ surface      & $C_{2v}$ &  0  &  $\varphi\sigma_z + \lambda_{R,x}q_z \sigma_x - \lambda_{R,z} q_x \sigma_z$\\
         \hline
         (2$\overline{1}$0) thin film    & $D_{2h}$ &   $\varphi q_y q_z \sigma_y$ &  $\varphi \sigma_z$ \\
         (2$\overline{1}$0) surface      & $C_{2v}$ &   $\varphi q_y q_z \sigma_y$ &  $\varphi \sigma_z + \lambda_{R,z}q_y \sigma_z - \lambda_{R,y} q_z \sigma_y$\\
         \hline
    \end{tabular}
    \label{tab:MnTeSummary}
\end{table*}

\section{Symmetry analysis for thin films and surfaces of MnTe}
MnTe is isostructural to CrSb. In the symmetry analysis, the main distinction is that the orientation of the N\'eel vector in MnTe lies within the xy-plane, specifically along the [120] axis, i.e., y-axis \cite{Lee2024}. Therefore, the symmetry analysis proceeds analogously to that for CrSb, and the results are summarized in Table \ref{tab:MnTeSummary}. For the cases involving the $(010)$ and the $(2\overline{1}0)$ plane, the altermagnetic OP $\varphi = \mathcal{O}_{(3x^2-y^2)yz,y}$ forms a one-dimensional irreducible representation, similar to all cases in CrSb. In contrast, for the cases involving the $(001)$ plane, the altermagnetic OP, corresponding to the N\'eel vector in the y-axis, coincides with one component of a two-dimensional irreducible representation of the point group $D_{3h}$ and $C_{3v}$ for the thin-film and the half-infinite geometry, respectively. We denote the two-component OP as $\vec{\varphi}_{xy} = (\varphi_x, \varphi_y)^T = \left(\mathcal{O}_{(3x^2-y^2)yz,x}, \mathcal{O}_{(3x^2-y^2)yz,y} \right)^T$, and we finally need to set $\varphi_x = 0$ in Table \ref{tab:MnTeSummary} for MnTe.

\section{Tight Binding model}

\begin{figure}[hbt]
    \centering
    \includegraphics[width=.5\linewidth]{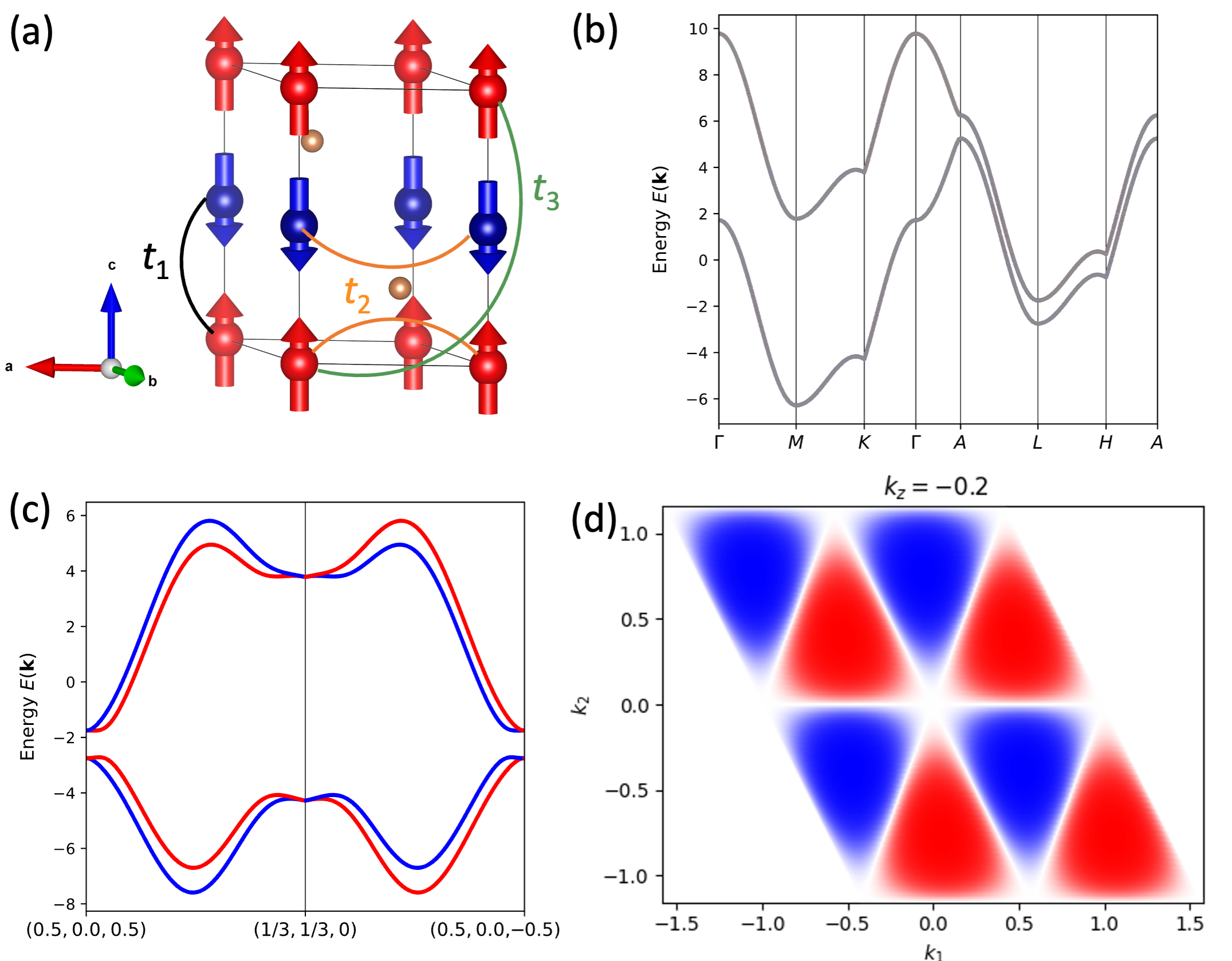}
    \caption{{\bf Tight-binding model of CrSb.} (a) Crystal structure of CrSb showing the different hopping paths used for building the tight-binding model. Electronic band structures calculated using the tight-binding Hamiltonian along (b) high-symmetry directions and (c) non-high-symmetry path in the Brillouin zone. (d) Momentum-resolved map of the spin splitting in the Brillouin zone for a constant $k_z$ = -0.2, illustrating the characteristic $g$-wave altermagnetic spin-splitting pattern.}
    \label{fig:Tb_S1}
\end{figure}

Tight-binding (TB) models provide a useful framework for analyzing the electronic band structure of bulk crystals and for studying how these properties evolve in the thin-film limit, such as in monolayers or few-layer systems. Here, we employ TB models to investigate thin films of a $g$-wave AM with different surface orientations. Our focus is on the orientation dependence of the electronic structure and the emergence of altermagnetic spin splitting or absence thereof, upon dimensionality reduction.
We exploit the the tight-binding model from Ref. \cite{PhysRevB.110.144412}, with minimal ingredients required to realize g-wave altermagnetic order being: (i) anisotropic hopping amplitudes along six-fold crystal directions and (ii) an on-site potential that distinguishes the magnetic sublattices. Together, these ingredients capture the essential symmetry breaking responsible for the characteristic spin splitting of altermagnets.

We analyze (001), (2$\overline{1}$0), and (010) oriented films, where our group-theoretical and simple surface scattering considerations already suggest that the spin splitting symmetry in thin films depends strongly on the chosen surface orientation. In particular, symmetry analysis indicates that two of these orientations ((001) and (010)) should host spin-degenerate bands, while one orientation should allow a higher-symmetry $d$-wave–like altermagnetic order to emerge in the thin-film geometry of CrSb type of altermagnets.
We make use of the python code \texttt{pythtb} package\cite{Cole_Python_Tight_Binding_2025} to implement TB model. To validate the correctness of the model, the altermagnetic band splitting was first reproduced for the bulk crystal structure. The relevant hopping pathways are illustrated in Fig. \ref{fig:Tb_S1} (a), with hopping amplitudes chosen as $t_1$, $t_2 = 0.5\,t_1$, $t_3 = \pm t_2$.
This parameter set for the representative model captures the characteristic g-wave altermagnetic spin splitting, observed in CrSb. The resulting electronic band structure and momentum-dependent spin splitting are shown in Fig. \ref{fig:Tb_S1} (b-d), demonstrating the successful reproduction of the expected altermagnetic features within the tight-binding framework.

\begin{figure}[h]
    \centering
    \includegraphics[width=0.5\linewidth]{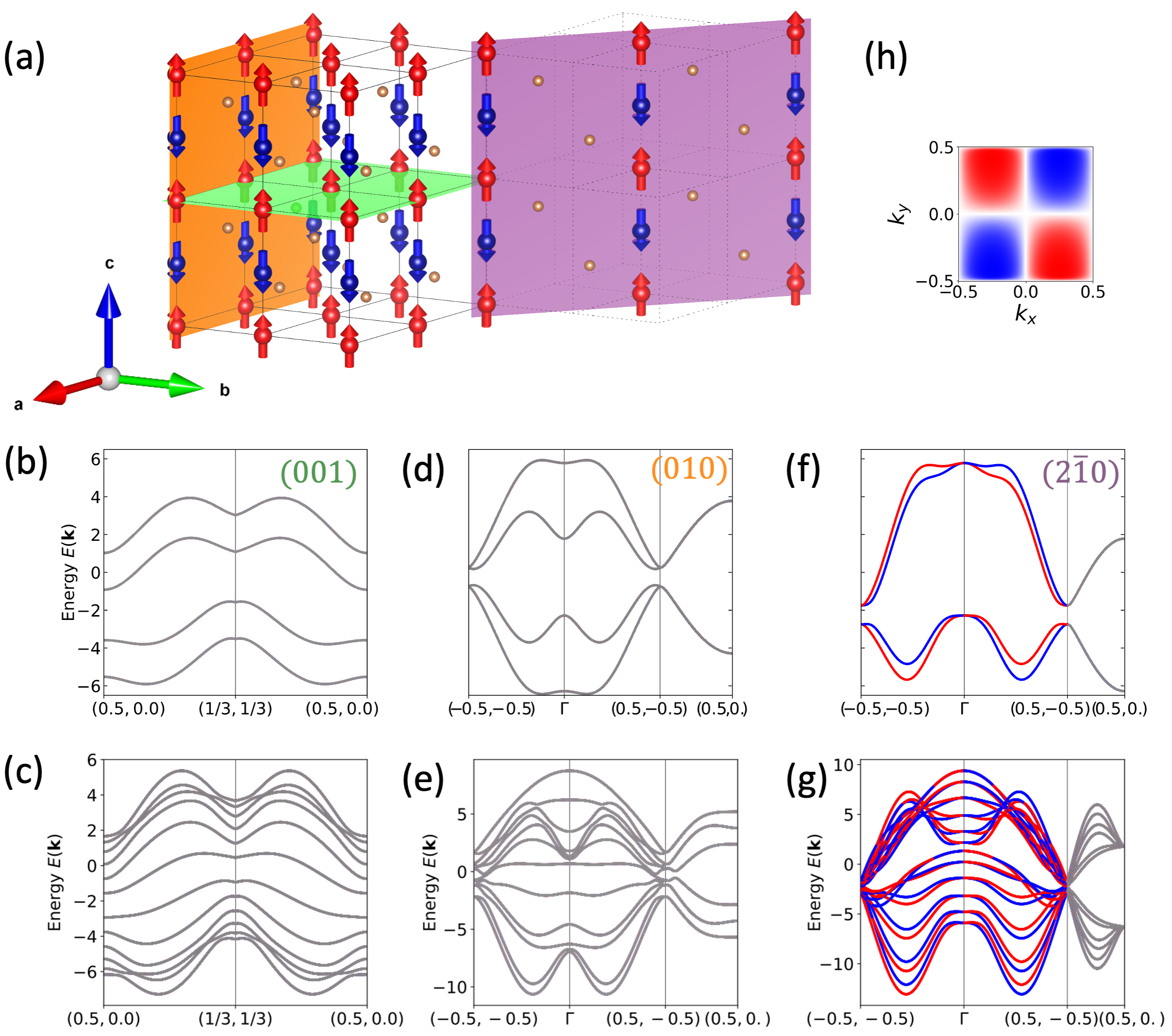}
    \caption{{\bf Tight-binding model of CrSb.} (a) Hexagonal crystal structure of CrSb, with three distinct crystallographic planes highlighted using different colors. Electronic band structures calculated from the tight-binding model for the (b-c) (001), (d-e) (010) and (f-g) (2$\overline{1}$0) plane. The lower panel for each plane represent electronic band structures calculated for films containing 12 orbitals/atoms, corresponding to 6 atomic layers.  (e) Mapping of spin splitting for the 2-D BZ for (2$\overline{1}$0) plane.}
    \label{fig:crsb}
\end{figure}

Next we consider the (001) oriented film, which typically coincides with a nodal plane of the bulk Brillouin zone for a $g$-wave AMs. Figure ~\ref{fig:crsb}(a) shows the different planes studies in the equivalent CrSb structure and corresponding band structure. The band calculations are presented for 1-2 monolayers for better clarity in Fig. ~\ref{fig:crsb} (b,d,f), while results for thicker slab are provided in  (c,e,g) panel. In agreement with symmetry considerations, the bands remain spin-degenerate for all film thicknesses for the (001) and for the (010) plane.
In contrast, the (2$\overline{1}$0) surface exhibits a clear manifestation of altermagnetic spin splitting. Remarkably, a pronounced splitting already appears in the monolayer limit and persists upon increasing the film thickness. This behavior is consistent with symmetry analysis, which identifies this orientation as allowing the altermagnetic order to be directly encoded in the surface electronic structure. A momentum-resolved mapping of the spin splitting across the 2-D BZ reveals a characteristic $d$-wave–like pattern with two nodal lines.

\section{Density-functional calculations}

The electronic structure calculations were performed within density-functional
theory using the full-potential linearized augmented plane-wave (FP-LAPW)
method as implemented in the WIEN2k package \cite{Blaha2020}.
The obtained electronic structure is in good agreement with previous
theoretical studies of CrSb \cite{Li2025,Ding2024,Reimers2024}.
Exchange-correlation effects were treated within the generalized-gradient
approximation (GGA) using the Perdew--Burke--Ernzerhof (PBE) functional.
Spin-orbit coupling was included using the second-variational method.
No on-site Hubbard correction ($+U$) was applied.

\begin{figure}[bh]
    \centering
    \includegraphics[width=0.2\linewidth]{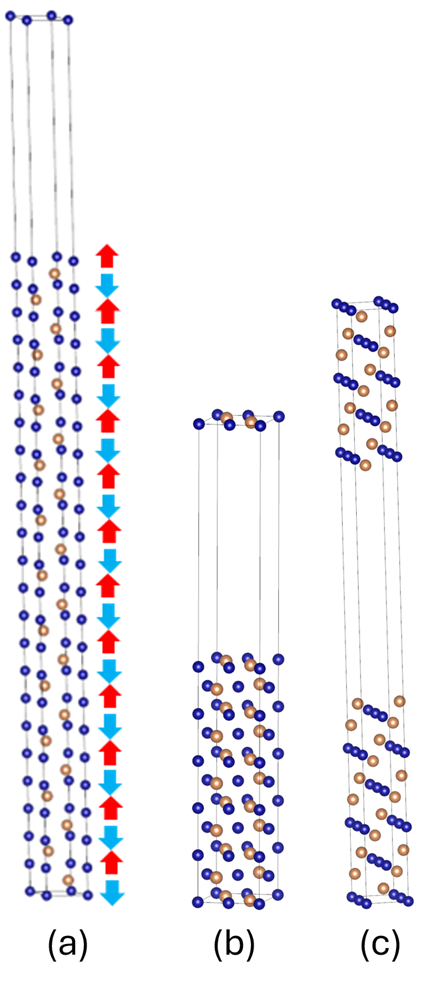}
\caption{
{\bf Crystal structures employed in the density-functional slab calculations.}
(a) Supercell used for the (001) surface calculation; red and blue arrows depict the Néel ordering of the magnetic moments on the Cr atoms.
(b) Supercell used for the (010) surface calculation.
(c) Supercell used for the (2$\overline{1}$0) surface calculation.
Vacuum regions of approximately 40 bohr were introduced between periodic slab images to suppress interactions between opposite surfaces.
Blue and orange spheres denote Cr and Sb atoms, respectively.
}
    \label{fig:Slabs}
\end{figure}

The calculations used the hexagonal NiAs-type CrSb structure with lattice parameters
$a=b=4.103$~\AA and
$c=5.463$~\AA \cite{Reimers2024}.
In particular, the calculated electronic structure of the (120) surface is
consistent with the ARPES measurements reported by Ding \textit{et al.}
\cite{Ding2024}. The muffin-tin radii were chosen as $R_{\mathrm{MT}}=2.5$ bohr for both Cr and Sb atoms, and the basis-set size was controlled by $R_{\mathrm{MT}}K_{\mathrm{max}}=7.5$.

To investigate the surface electronic structure, explicit slab supercells
were constructed for the (001), (100), and (120) surface orientations
(Fig.~\ref{fig:Slabs}).
A vacuum region of approximately 40 bohr was introduced between
periodic slab images to suppress artificial interactions between
opposite surfaces.
The atomic positions were kept fixed at their bulk crystallographic
values, i.e. no structural relaxation was performed.
Spin-orbit coupling was included in all slab calculations.

Brillouin-zone integrations for the bulk calculations were performed
using a $31\times61\times121$ $k$-point mesh.
The projected bulk band structures shown in the main text were obtained
by projecting the three-dimensional bulk electronic structure onto the
corresponding surface Brillouin zones and comparing them directly with
the slab calculations.

\bibliography{refs}